\documentclass[12pt,preprint]{aastex}

\shorttitle{BEHAVIOR OF PITCH ANGLE AS A FUNCTION OF WAVELENGTH}
\shortauthors{Mart\'{\i}nez-Garc\'ia et al.}

\begin{document}

\title{THE BEHAVIOR OF THE PITCH ANGLE OF SPIRAL ARMS DEPENDING ON OPTICAL WAVELENGTH}

\author{Eric E. Mart\'inez-Garc\'ia\altaffilmark{1}, 
Iv\^anio Puerari\altaffilmark{1}, 
F. F. Rosales-Ortega\altaffilmark{1},
Rosa A. Gonz\'alez-L\'opezlira\altaffilmark{2},
Isaura Fuentes-Carrera\altaffilmark{3},  and
A. Luna\altaffilmark{1}}

\affil{1 Instituto Nacional de Astrof\'isica, \'Optica y Electr\'onica (INAOE), Aptdo. Postal 51 y 216, 72000 Puebla, Pue., M\'exico.}
\affil{2 Centro de Radioastronom\'ia y Astrof\'isica, UNAM, Campus Morelia, Michoac\'an, M\'exico, C.P. 58089}
\affil{3 Escuela Superior de F\'isica y Matem\'aticas, Instituto Polit\'ecnico Nacional, U. P. Adolfo L\'opez Mateos, Zacatenco, 07730 M\'exico, D. F., M\'exico.} 

\email{ericmartinez@inaoep.mx}

\begin{abstract}

Based on integral field spectroscopy data from the CALIFA survey,
we investigate the possible dependence of spiral arm pitch angle with optical wavelength.
For three of the five studied objects, the pitch angle gradually increases at longer wavelengths.
This is not the case for two objects where the pitch angle remains constant.
This result is confirmed by the analysis of SDSS data.
We discuss the possible physical mechanisms to explain this phenomenon, as well as the implications of the results.

\end{abstract}

\keywords{ galaxies: fundamental parameters --- galaxies: spiral --- galaxies: structure --- methods: observational
--- methods: numerical }

\section{Introduction}

Spiral structure in disk galaxies has been a matter of study for decades.
The origin of spiral arms of the grand-design type is still a subject of debate,
although the density wave (DW) theory~\citep{lin64,ber89} has provided an adequate explanation in the past decades.
In the case of non-barred (or weakly barred) objects, recent numerical simulations~\citep[e.g.,][]{sel11,bab13,don13,roc13}
show transient short-lived spiral arms, different from those expected by DW theory.
Nevertheless, long-lived structures have been obtained by~\citet[][see also associated articles, and references therein]{zha98}.
For barred objects, new theories~\citep[e.g., {\it{manifolds}},][]{rom06,vogs06,pat06,tso08,atha09,atha10}
have emerged that propose different interpretations of spiral arm dynamics~\citep[see also][]{mar12}.

Spiral arms can be observed up to a redshift of $z\sim2$~\citep{elm14}, when the universe was only $\sim3.4$ Gyr old.
From these epochs until the recent era, the morphology of the spirals provides
important clues to determine the physical mechanisms that generate
spiral structure, and the onset of young star formation that accompanies it.
Among the morphological parameters, the pitch angle\footnote{The pitch angle is defined as
the angle between a tangent to the spiral arm at a certain point and a circle,
whose center coincides with the galaxy's, crossing the same point.}
is one of the easiest to determine without the need to assume a prior model. 
When different wavebands are involved in studies referent to pitch angles,
some authors propose using a wide range of wavelengths~\citep[e.g.,][]{sei06,dav12}.
Nonetheless, indications of pitch angles being tighter in bluer than in
redder colors have also been obtained~\citep{gro98,mar12,mar13}.

This research aims to analyze data from integral field spectroscopy (IFS),
in order to explore the behavior of pitch angle with wavelength ($\lambda$).
It also has the purpose of discussing the physical
processes that could determine pitch angle as a function of $\lambda$.

\section{Data sample}~\label{data_sample}

Our sample consists of 5 nearly face-on spiral galaxies (see table~\ref{tbl-1}), with integral field spectroscopy (IFS)
data taken from the Calar Alto Legacy Integral Field Area survey~\citep[CALIFA,][]{san12}.
The CALIFA survey uses the Potsdam Multi-Aperture Spectrophotometer, 
and fiber pack~\citep[PMAS/PPak;][]{rot05,kel06},
mounted on the 3.5 meter telescope at the Calar Alto Observatory. We use datacubes (three-dimensional data) from the first
public data release~\citep[DR1,][]{hus13}, and the spectral setup V500.
This setup covers the wavelength range 3749-7501~\AA\AA, with a
wavelength sampling of 2~\AA\AA~per pixel.
Due to internal vignetting within the spectrograph,
the fraction of valid (or useful) pixels differs from 100\%, descending gradually to $\sim70$\%
for $4240>\lambda>7140$~\AA\AA.
The plate scale is $1\arcsec$~pixel$^{-1}$, and the point spread function full width at half max (PSF FWHM) is $3\farcs7$.

The objects were selected by visually inspecting the datacubes;
we searched for objects with well defined spiral arms and inclinations to the line of sight smaller than 60$\degr$.
Our initial selection consisted of 10 objects, of which we excluded five
(NGC~4210, NGC~5378, NGC~7819, UGC~8781, and UGC~9476).
The exclusion was due to the failure to conduct a proper analysis (see also section~\ref{results})
of the pitch angles, owing to insufficient spatial resolution and the small 
field-of-view (FOV) provided by the CALIFA instrument.

To complement our study we also analyze images from the Eighth Sloan Digital Sky
Survey Data Release~\citep[SDSS DR8;][]{aih11} at the
$u$, $g$, $r$, $i$, and $z$ bands. These have a PSF FWHM of $\sim1\farcs5$.
H$\alpha$ emission line maps were also considered.
These were obtained from the IFS datacubes using the PPAK IFS Nearby Galaxy Survey
software~\citep[{\sc PINGSoft};][]{ros11}, by applying an H$\alpha$ narrow band
20\AA\AA~width, centered on the H$\alpha$ emission line of each object, and subsequently
subtracting a local adjacent continuum.

Within our sample, only NGC~776 and NGC~5406 present a bar structure in their disk (SAB type),
while the rest of the objects are unbarred (see also figure~\ref{FFT_annuli}).
The NASA/IPAC Extragalactic Database (NED)\footnote{http://ned.ipac.caltech.edu/} homogenized hierarchy parameter indicates
that NGC~776, NGC~4185, and NGC~5406 are in a radial velocity-based grouping~\citep{mat98};
NGC~2916 is an isolated galaxy~\citep[6 Mpc isolation;][]{vet86}; and UGC~7012 is
a pair member~\citep{kar72}.

\section{Analysis}~\label{analysis}

The projection parameters (inclination and position angle)
were obtained from the SDSS $i$-band images by fitting ellipses, with IRAF,\footnote{
IRAF is distributed by the National Optical Astronomy Observatories,
which are operated by the Association of Universities for Research in
Astronomy, Inc., under cooperative agreement with the National Science
Foundation.}
to the outer isophotes of the disks. We assume that these parameters are roughly
constant with wavelength. We then use these parameters (see table~\ref{tbl-1})
to obtain the deprojected datacubes, and images, for each object.

The pitch angles were measured through
the two-dimensional Fourier transform (2DFT) method~\citep[e.g.,][]{kal75,con82,con88,pue92,sar94,dav12,sav12}.
The coordinates (Cartesian, $x$ vs. $y$) of each pixel, in each slice\footnote{
We define ``slice'' as a section of the datacube with constant wavelength.
In this manner we have 1877 slices for each datacube in the wavelength
range 3749-7501~\AA\AA.}
of the deprojected datacubes were transformed to $\theta$ vs. $u=\ln R$ coordinates.
In this way, a ($\theta, u$) datacube was obtained
for each object. These ($\theta, u$) datacubes were then analyzed with the 2DFT method,
by applying (for each slice, respectively):

\begin{equation}~\label{eq2DFT}
A(m,p)=\int_{u_{\mathrm{min}}}^{u_{\mathrm{max}}} \int_{-\pi}^{\pi}
I(\theta,u)e^{-i(m\theta+pu)}d\theta du,
\end{equation}

\noindent where $I(\theta,u)$ represents the mean intensity of radiation at the $\theta,u$ coordinates,
and $u_{\mathrm{min}},u_{\mathrm{max}}$ are the radial limits where the {\it{average}} pitch 
angle is obtained.\footnote{A new method recently proposed by Puerari, Elmegreen, \& Block (2014, submitted)
involves the measurements of pitch angles locally, confining calculations to a window in the $\theta$ vs. $u$ plane.}
The use of equation~\ref{eq2DFT} implies the representation of 
the observed spiral arms with a set of logarithmic spirals, in the same fashion that
a signal can be represented by a set of sine and cosine functions.
The outcome is the $A(m,p)$ matrix, for modes $m=0, 1, 2, 3, 4, ...$, having
$m=2$ for two-armed spirals. For a certain mode $m$, a spectrum is obtained
for several $p$ frequencies, and the maximum in this spectrum, $p_{\mathrm{max}}$,
is related to the pitch angle, $P$, as

\begin{equation}~\label{eqP}
 \tan{P} = -m/p_{\mathrm{max}},
\end{equation}

\noindent where positive or negative $P$ values can be obtained,
depending on whether the object has a ``S'' or ``Z'' view on the sky, respectively.

The maximum in the $A(m,p)$ spectrum is certainly affected by the choice of the $u_{\mathrm{min}},u_{\mathrm{max}}$
limits, especially the $u_{\mathrm{min}}$ value~\citep{dav12}. Nevertheless,
for the purpose of analyzing a possible variation of the pitch angle with wavelength,
it is important to fix the radial limits in order to avoid biases introduced by
variations of the pitch angle with radius, which have been noted~\citep[e.g.,][]{sav13,pue14}.
Bar or bulge distortions, inside the $u_{\mathrm{min}}-u_{\mathrm{max}}$ region,
can also give a maximum in the $A(m,p)$ spectrum (for $p\sim0$, i.e., $P\sim90\degr$) which is not associated to the spiral arms.
For this reason each spectrum was systematically inspected to ensure that the maximum, {{\it{absolute}} or {\it{relative}}, belongs
to the spiral arms.

The spirals in the CALIFA datacubes, as well as the SDSS and H$\alpha$ images,
were analyzed within the radial regions listed in table~\ref{tbl-2} (see also figure~\ref{FFT_annuli}).
On average, these regions cover a fraction of 0.1-0.3$R_{25}$ in the disks.
We adopt $m=2$ for the five objects in our sample.

\subsection{Error analysis}

The error associated to the SDSS and H$\alpha$ pitch angles consists of two parts, added in quadrature.
The first one corresponds to the numerical error of the method,
which is simply the separation between $p_{\mathrm{max}}$ and
the adjacent $p$ frequencies in the $A(m,p)$ matrix, with
dimensions $N_{m} \times N_{p}$.
This separation is given by discrete and equal steps, $\Delta p$,
which are determined in our numerical code\footnote{
The code is based on the two-dimensional fast Fourier
transform routine~{\tt{FOURN}}~\citep{pre92}.}
by:
 
\begin{equation}~\label{delp}
 \Delta p = \frac{2\pi}{N_{p} \left(\frac{u_{\rm ext}}{R_{\rm pix}}\right) }.
\end{equation}

The quantity $u_{\rm ext} = \ln R_{\rm ext}$ is related to the
pixel in the Cartesian image with the largest radial extension, measured from the center of the
object (commonly located in the center of the image).
As explained before, we first transform all the pixels in our input Cartesian image
to a ($\theta, u$) array, with dimensions $N_{m} \times R_{\rm pix}$,
where $R_{\rm pix} < N_{p}$. The elements in the array outside the region
$u_{\mathrm{min}}-u_{\mathrm{max}}$ are set to zero.
After various tests we found that optimal results are obtained with 
$N_{m} \times N_{p} = 128 \times 2048$, and $R_{\rm pix} = 256$.
An image of dimensions 2048 (columns) $\times$ 2048 (lines)
would have $u_{\rm ext} \sim 7.3$, and $\Delta p \sim 0.11$.
This is an important distinction in comparison with other numerical codes,
where $\Delta p$ is fixed to 0.25~\citep{sar94,sei06,dav12}. 

The second part of the error is related to the flux errors in each pixel.
Flux errors were assessed through numerical Monte Carlo simulations.
For each pixel and at each SDSS passband, the fluxes were assigned 
a Gaussian probability distribution with
$\sigma=0.04$ magnitudes.
For the H$\alpha$ images, on the other hand, we use the corresponding flux error map.
At each band, we derive the pitch angle 100 times, 
and compute the corresponding $1\sigma$ standard deviation for $P$.
We find that in most cases this source of error is not significant. 

The error associated to the CALIFA pitch angles
is already taken into account by the differences in the 
measurements for adjacent slices (or wavelengths).
In our analysis we did not detect the need to increase the S/N ratio
for each slice, which can be accomplished by integrating in $\Delta\lambda$.


\begin{deluxetable}{ccccc}
\tabletypesize{\scriptsize}
\tablecaption{Data parameters~\label{tbl-1}}
\tablewidth{0pt}
\tablehead{
\colhead{Object} &
\colhead{Type}   &
\colhead{P.A.}   &
\colhead{Incl.}  &
\colhead{Dist.}
}
\startdata

NGC~~776    & SAB(rs)b   &  ~30.4     &     29    &  65.5   \\
NGC~2916    & SA(rs)b    &  ~24.6     &     46    &  56.0   \\
NGC~4185    & Sbc        &  166.5     &     49    &  61.0   \\
NGC~5406    & SAB(rs)bc  &  110.3     &     44    &  79.0   \\
UGC~7012    & Scd        &  ~13.7     &     55    &  49.4   \\

\enddata

\tablecomments{
Column 2: morphological type~\citep[RC3,][]{dva91}.
Columns 3 and 4: position angle and inclination in degrees.
Column 5: Hubble flow distance, in Mpc, from the NASA/IPAC extragalactic database (Virgo + Great Attractor + Shapley Supercluster);
$H_{0}=73$ km s$^{-1}$ Mpc$^{-1}$, $\Omega_{\mathrm{matter}}=0.27$, and $\Omega_{\mathrm{vacuum}}=0.73$.
}

\end{deluxetable}


\section{Results}~\label{results}

The pitch angles obtained for the SDSS and H$\alpha$ data are tabulated in table~\ref{tbl-2}.
These values are also shown in figure~\ref{Pangles} with
triangle symbols. In these figures we also plot the pitch angle values, as a function of wavelength,
for the CALIFA datacubes with cross symbols. For reference, we added to these figures a dotted line that indicates
the pitch angle obtained for the SDSS $z$-band, and a solid line which represents the
second-order best-fit polynomial to the CALIFA's data.
The fits to the polynomials are summarized as follows:

\begin{itemize}

\item[] NGC~ 776:~$P(\lambda)= -6.129 + (4.079\times10^{-4})\lambda - (4.006\times10^{-8})\lambda^{2}$;~~rms=0.22$\degr$
\item[] NGC~2916:~$P(\lambda)= -8.543 + (8.938\times10^{-3})\lambda - (4.804\times10^{-7})\lambda^{2}$;~~rms=2.89$\degr$
\item[] NGC~4185:~$P(\lambda)=  5.997 + (8.835\times10^{-4})\lambda - (5.112\times10^{-9})\lambda^{2}$;~~rms=1.00$\degr$
\item[] NGC~5406:~$P(\lambda)= 13.409 + (5.364\times10^{-3})\lambda - (4.187\times10^{-7})\lambda^{2}$;~~rms=0.95$\degr$
\item[] UGC~7012:~$P(\lambda)= 19.721 - (1.523\times10^{-6})\lambda - (3.927\times10^{-9})\lambda^{2}$;~~rms=0.85$\degr$

\end{itemize}

\noindent $P$ is in degrees ($\degr$) and $\lambda$, in angstroms (\AA\AA).

From the figures it can be appreciated that for three of the objects,
NGC~2916, NGC~4185, and NGC~5406, the pitch angle tends to gradually increase with wavelength, while for
NGC~776 and UGC~7012 the pitch angle has a fairly constant value.

In the case of UGC~7012, we notice that the mean pitch angle derived from CALIFA is $\sim2\degr$ lower
when compared to the one from SDSS. We attribute this difference to the different spatial resolutions of the two instruments.
To a lesser extent, something similar occurs to NGC~776, and NGC~5406.
This indicates that the resolution of the image can also affect the measurements of the pitch angles.
For NGC~2916, the effects of vignetting (see section~\ref{data_sample}) can
be appreciated for $3850\ga\lambda\ga7250$~\AA\AA, where the measurements of the pitch angles are more scattered.

For the SDSS data, the second-order best-fit polynomials (not plotted in figure~\ref{Pangles}) are:

\begin{itemize}

\item[] NGC~ 776:~$P(\lambda)= -5.911 + (8.907\times10^{-5})\lambda - (7.735\times10^{-9})\lambda^{2}$;~~rms=0.03$\degr$
\item[] NGC~2916:~$P(\lambda)=  8.967 + (4.417\times10^{-3})\lambda - (2.114\times10^{-7})\lambda^{2}$;~~rms=0.48$\degr$
\item[] NGC~4185:~$P(\lambda)=  1.469 + (2.743\times10^{-3})\lambda - (1.704\times10^{-7})\lambda^{2}$;~~rms=0.04$\degr$
\item[] NGC~5406:~$P(\lambda)= 14.839 + (3.555\times10^{-3})\lambda - (2.162\times10^{-7})\lambda^{2}$;~~rms=0.05$\degr$
\item[] UGC~7012:~$P(\lambda)= 17.455 + (1.143\times10^{-3})\lambda - (6.919\times10^{-8})\lambda^{2}$;~~rms=0.31$\degr$

\end{itemize}

\noindent If a constant is fitted to all five SDSS data points, for instance $P(\lambda)=z$-band pitch angle,
we get rms=0.07$\degr$, 5.32$\degr$, 1.68$\degr$, 2.39$\degr$, and 0.96$\degr$; for NGC~776, NGC~2916, NGC~4185, NGC~5406, and UGC~7012, respectively. Comparing these with the rms values of the second-order polynomials (SDSS),
it can be argued quantitatively that only for NGC~776 and UGC~7012
a constant pitch angle can be adopted for the respective wavelength range.


\begin{deluxetable}{clcccccc}
\tabletypesize{\scriptsize}
\tablecaption{Pitch angles~\label{tbl-2}}
\tablewidth{0pt}
\tablehead{

\colhead{Object} &
\colhead{$\Delta{R}$} &
\colhead{$u$-band~(\degr)} &
\colhead{$g$-band~(\degr)} &
\colhead{$r$-band~(\degr)} &
\colhead{$i$-band~(\degr)} &
\colhead{$z$-band~(\degr)} &
\colhead{H$\alpha$~(\degr)}
}

\startdata

NGC~~776    & (16.6-34.8$\arcsec$)     & -5.67($\pm$0.06)  & -5.71($\pm$0.06)  & -5.63($\pm$0.06) & -5.67($\pm$0.06) & -5.74($\pm$0.06) & -5.87($\pm$0.06) \\~\vspace{1mm}
~~          & (0.16-0.33$R_{25}$) \\
NGC~2916    & (15.4-41.2$\arcsec$)     & 22.40($\pm$0.55)  & 24.16($\pm$0.90)  & 28.57($\pm$1.19) & 30.39($\pm$1.34) & 31.38($\pm$1.42) & 21.23($\pm$0.69) \\~\vspace{1mm}
~~          & (0.10-0.28$R_{25}$) \\
NGC~4185    & (17.8-36.8$\arcsec$)     & ~9.07($\pm$0.13)  & 10.58($\pm$0.18)  & 11.85($\pm$0.22) & 12.52($\pm$0.25) & 12.35($\pm$0.24) & ~9.81($\pm$0.15) \\~\vspace{1mm}
~~          & (0.11-0.23$R_{25}$) \\
NGC~5406    & (16.6-32.9$\arcsec$)     & 24.70($\pm$0.66)  & 26.84($\pm$0.77)  & 28.47($\pm$0.86) & 29.35($\pm$0.91) & 29.35($\pm$0.91) & 27.16($\pm$1.11) \\~\vspace{1mm}
~~          & (0.14-0.28$R_{25}$) \\
UGC~7012    & (5.5-15.8$\arcsec$)     & 20.37($\pm$0.45)  & 21.80($\pm$0.73)  & 21.80($\pm$0.73) & 21.80($\pm$0.73) & 22.32($\pm$0.74) & 20.00($\pm$0.59) \\~\vspace{1mm}
~~          & (0.06-0.16$R_{25}$) \\

\enddata

\tablecomments{
Column 2: radial regions, ($R_{\mathrm{min}}$-$R_{\mathrm{max}}$) in arcseconds, 
and as a fraction of $R_{25}$ (RC3, $B$-band),
where the average pitch angles are measured in SDSS and CALIFA data.
Columns 3-7: SDSS average pitch angles, in degrees, for the $u$, $g$, $r$, $i$ and $z$-bands.
Column 8: average pitch angle, in degrees, from H$\alpha$ emission maps.
}

\end{deluxetable}


\section{Discussion}

The fact that the spiral arms pitch angle tends to increase gradually with wavelength
indicates that different stellar populations are spatially located in an ordered manner.
While the old (red) stellar populations lie in loose spirals, the young (blue) stellar populations belong to tighter spirals.
We will discuss four possible physical mechanisms to explain this behavior.

The most obvious would be that extinction is affecting young stellar objects near the dust lanes,
on the concave side of the spiral arms. The problem would be to explain the systematic ordering of the dust
to produce the observed effect; dust is in general not observed to 
be distributed in a symmetrical manner near the spiral arms.

For a Doppler shift to cause the observed effect (i.e., the light of stars with similar colors being wavelength shifted 
from 4000 to 6000\AA\AA~due to the velocities at which they drift out of the spiral arms), 
the speeds required would be of the order of 0.5$c$. Hence, this mechanism is not really physically possible. 

Another explanation comes naturally from azimuthal age/color gradients across spiral arms~\citep[e.g.,][]{gon96,mar09,mar13}.
These gradients are the product of the triggering of star formation as a consequence of large-scale spiral shocks~\citep{rob69,git04}.
The oldest stars, that generate the required gravitational potential, would be located in the vicinity of the shock location.
Downstream this oldest population (following the gas flow direction), the youngest stars would be located
near active star-forming regions. These young stars will then gradually age, as they leave the place where they were born, 
and produce a gradient toward the red in the opposite direction.
In this way we will have the oldest population, followed by the youngest stars, and then a gradually aging population
downstream the shock.\footnote{
\citet{gon96} introduced a photometric index, $Q(rJgi)$, that effectively traces an age gradient from young to relatively
old populations. This index is mostly unaffected by the underlying oldest population,
regardless of its distribution near the spiral arms~\citep{mar09}.}
This scenario naturally develops loose spirals for the red populations, and then tighter spirals for the blue ones.
The pitch angles obtained for the H$\alpha$ emission support this line of reasoning,
since they tend to have similar values to the blue-band pitch angles, and not to the red-band ones.
A similar effect has been recovered in the hydrodynamic simulations of~\citet{kim14},
who find that the pitch angle of gaseous arms is
smaller than that of stellar arms~\citep[see also][]{git04,pat94}.
This is also supported by~\citet{sei02}, who used a combination of H$\alpha$ and $K$-band imaging,
for 20 spiral galaxies, and showed that there is enhanced star formation in the vicinity of spiral arms.
Under the proposed scenario, the ``blue spiral'' will intersect the ``red spiral'' near the corotation radius ($R_{\rm CR}$).
This would indicate that the angular speed of the spiral pattern ($\Omega_{p}$) can be assumed to be constant
for all radii~\citep{mar14}, at least in the object's current stage of evolution.

An earlier prediction by~\citet{hoz03} indicates a similar behavior, i.e.,
loose spirals for the red populations, and then tighter spirals for the blue ones.
In this case the physical mechanism is related to the difference in velocity dispersion between young
and old stars~\citep[see also][]{atha10,mar12}. However, this mechanism does not explain the similarity between the
pitch angles in H$\alpha$ and the blue bands.

To understand why some objects have constant pitch angles and others do not,
a statistical study is needed with a much larger sample. 
With this very small sample, neither distance nor inclination to the line of sight seem to be factors. 
Nonetheless, from a sample of 19 non-barred (or weakly barred) spirals,~\citet{mar13} have already shown
that~$\sim50\%$ of these objects show signs of azimuthal age/color gradients across their spiral arms.
It is also interesting to note that the variation of pitch angle with wavelength is 
most noticeable for NGC~2916, the only object cataloged as {\it{isolated}} in our sample.
Differences in the integrated values of the star formation rate (SFR),
and/or specific star formation rate (SFR per unit galaxy stellar mass, SSFR),
could account for the presence or absence of correlation between spiral arm pitch angle and wavelength.
In table~\ref{tbl-3} we list the SFRs, and SSFRs, for our sample.
The SFRs ​​show no clear trend, however the SSFRs take smaller values ​​for galaxies with a pitch angle
that depends on wavelength.


\begin{deluxetable}{cccc}
\tabletypesize{\scriptsize}
\tablecaption{Star formation rates~\label{tbl-3}}
\tablewidth{0pt}
\tablehead{
\colhead{Object} &
\colhead{log SFR (H$\alpha$)}   &
\colhead{log $M_{*}$}   &
\colhead{log SSFR}
}
\startdata

NGC~~776    &   ~0.52           &     10.65      &     -10.13  \\
NGC~2916    &   ~0.28           &     10.55      &     -10.27  \\
NGC~4185    &   -0.07           &     10.75      &     -10.82  \\
NGC~5406    &   ~0.33           &     11.05      &     -10.72  \\
UGC~7012    &   -0.18           &    ~9.15       &    ~-9.33   \\

\enddata

\tablecomments{
Column 2: logarithm of the integrated value of the star formation rate (SFR in $M_{\sun}$ yr$^{-1}$);
derived from the H$\alpha$ maps~\citep{ken98}.
Column 3: logarithm of the total stellar mass (in $M_{\sun}$) of the galaxy.
Column 4: logarithm of the integrated value of the specific star formation rate (SSFR, yr$^{-1}$).
}

\end{deluxetable}


\section{Conclusions}

Using IFS CALIFA survey data and SDSS images of five spiral galaxies, 
we have found that in two of them the arm pitch angle ($P$) stays constant with wavelength ($\lambda$), 
at least in the optical. The remaining three, however,
show clear signs of a dependence of $P$ with $\lambda$.
The analysis, via Fourier techniques, indicates a gradual increase of $P$ with increasing $\lambda$.
Although based on a small number of objects,
these results add statistical weight to the previous findings
of~\citet{gro98},~\citet{mar12}, and~\citet{mar13}.
This notwithstanding, the relationships found between spiral arm pitch angle and supermassive black hole mass
by~\citet{sei08}, and~\citet{ber13}
are likely to hold, given the larger errors of these studies.
Even so, it would be important to consider the correlation between wavelength and pitch angle for future works.

\acknowledgments

We thank an anonymous referee for valuable comments. 
EMG acknowledges the fellowship from the Mexican institution CONACYT, for a postdoctoral stay at INAOE.
IP acknowledges support from CONACYT.

This study makes use of the data provided by the Calar Alto Legacy
Integral Field Area (CALIFA) survey (http://califa.caha.es/).
Based on observations collected at the Centro Astron\'omico Hispano
Alem\'an (CAHA) at Calar Alto, operated jointly by the Max-Planck-
Institut f\"{u}r Astronomie (MPIA) and the Instituto de Astrof\'isica de
Andaluc\'ia (CSIC).

This research has made use of the NASA/IPAC Extragalactic Database (NED) which is operated
by the Jet Propulsion Laboratory, California Institute of Technology,
under contract with the National Aeronautics and Space Administration.


\begin{figure}
\centering
\epsscale{1.0}
\plotone{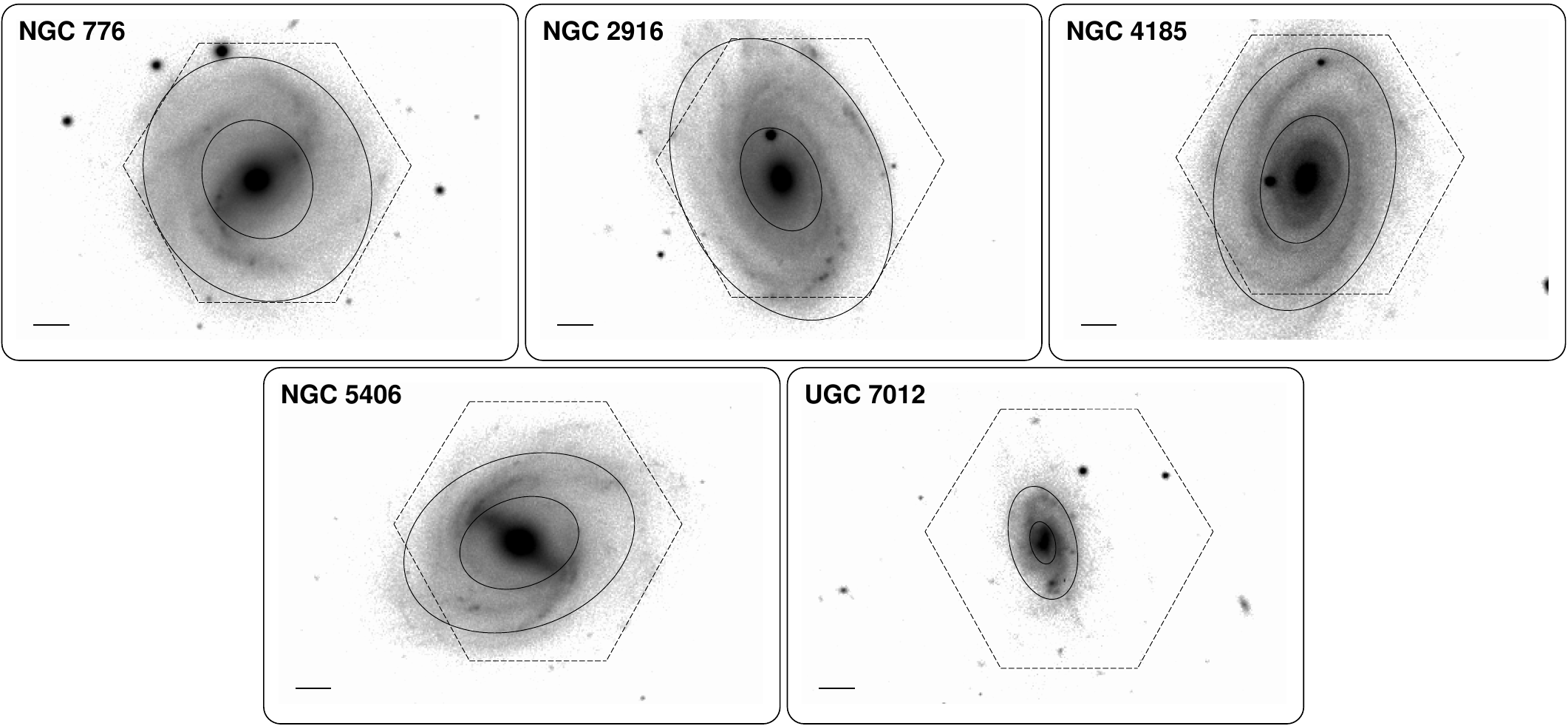}
\caption[f1]{SDSS $i$-band mosaics (not deprojected) for our sample. North is up, east to the left.
For each object, the radial regions, $R_{\mathrm{min}}$-$R_{\mathrm{max}}$
(cf. table~\ref{tbl-2}), where the 2DFT method was applied (see section~\ref{analysis}) are shown with solid lines.
In each frame the horizontal line in the lower left corner, represents $10\arcsec$. 
The dotted-line hexagon marks the CALIFA FOV. The gray scale is logarithmic.
~\label{FFT_annuli}}
\end{figure}

\begin{figure}
\centering
\includegraphics[scale=0.6,angle=270]{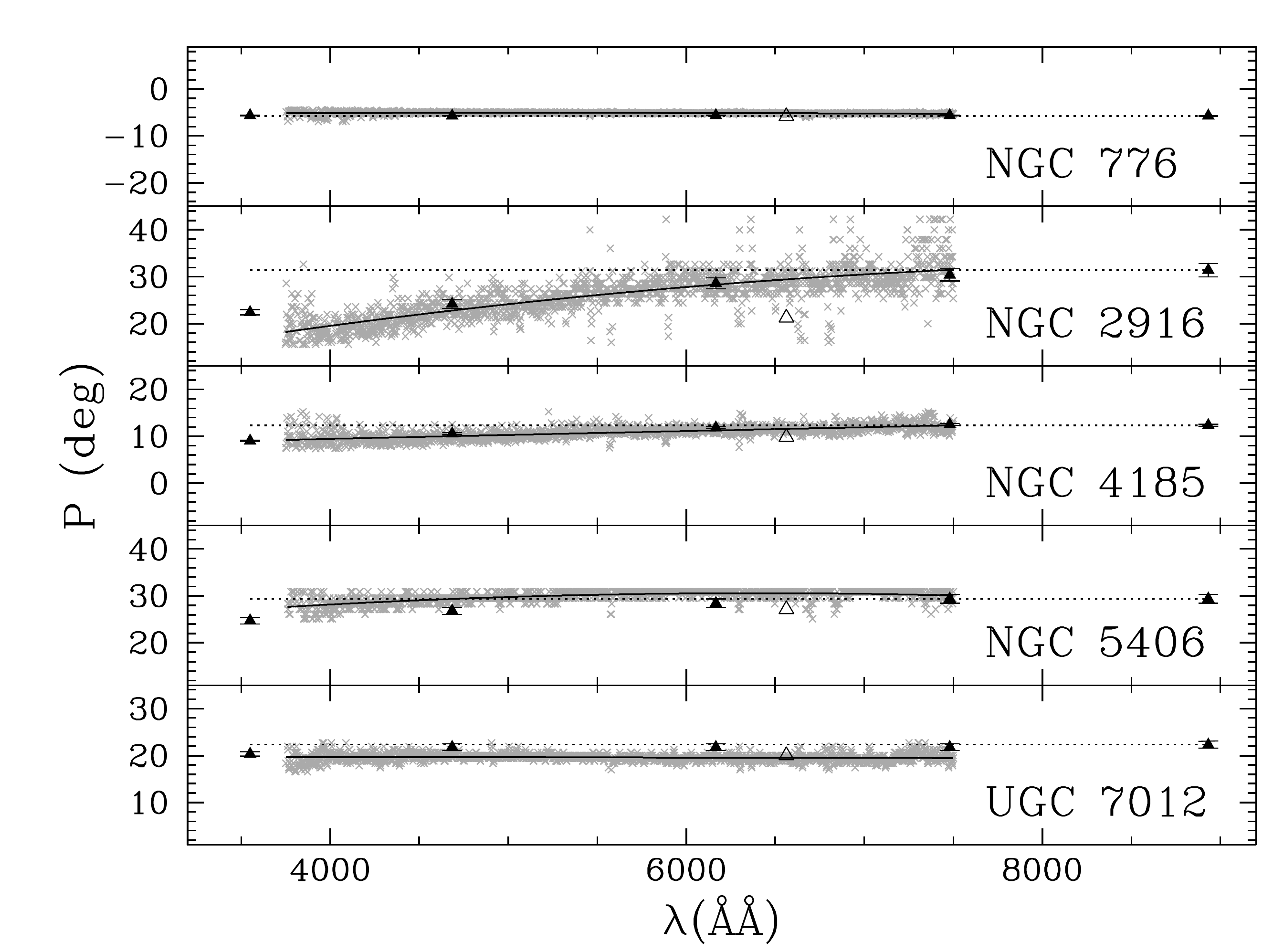}
\caption[f2]{Pitch angle values $P$ vs. wavelength $\lambda$.
{\it Solid black triangles with error bars:} SDSS; {\it empty triangles:} H$\alpha$; {\it gray crosses:} CALIFA.
Notice the similarity between the H$\alpha$ and bluest pitch angles.
The dotted line represents the SDSS $z$-band pitch angle value;
the solid line is the second-order best-fit polynomial to the CALIFA data.
~\label{Pangles}}
\end{figure}

\end{document}